\documentclass[a4paper]{VulcanTechReport}

\usepackage{cite}
\usepackage{tikz}
\usetikzlibrary{shapes.geometric,arrows.meta,positioning,calc,backgrounds}
\usepackage{amsmath}
\usepackage{amssymb}
\usepackage{adjustbox}
\usepackage{booktabs}
\usepackage{colortbl}
\usepackage{listings}

\definecolor{vulcanyellow}{RGB}{255,202,58}
\definecolor{ttcolor}{HTML}{1A1A1A}
\definecolor{codebg}{HTML}{F0F0F0}
\hypersetup{urlcolor=ttcolor}

\usepackage[most]{tcolorbox}
\newtcbox{\code}{on line,
  boxrule=0pt, boxsep=0pt,
  top=2pt, bottom=2pt, left=3pt, right=3pt,
  arc=3pt,
  colback=codebg, coltext=ttcolor,
  fontupper=\ttfamily}
\newcommand{\icode}[1]{{\color{ttcolor}\ttfamily #1}}

\lstdefinestyle{yamlstyle}{
  basicstyle=\ttfamily\scriptsize,
  frame=none,
  numbers=none,
  showstringspaces=false,
  breaklines=true,
  xleftmargin=0pt,
  xrightmargin=0pt,
  aboveskip=0pt,
  belowskip=0pt,
}
\newtcolorbox{codebox}{
  colback=codebg, colframe=codebg, boxrule=0pt, arc=0pt,
  left=4pt, right=4pt, top=3pt, bottom=3pt,
  before skip=4pt, after skip=4pt,
}

\reporttitle{IPI-proxy: An Intercepting Proxy for Red-Teaming Web-Browsing AI Agents Against Indirect Prompt Injection}
\reportsubtitle{Vulcan Research, AIFT}
\reportauthors{Chia-Pei (Janet) Chen, Kentaroh Toyoda, Anita Lai, and Alex Leung}
\reportdate{May 2026}

\leftheadercontent{IPI-proxy}
\rightheadercontent{\includegraphics[width=3cm]{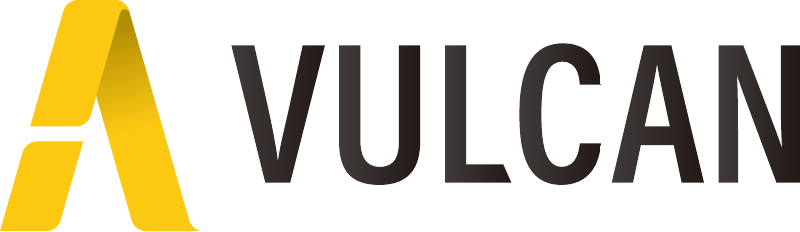}}

\begin{document}


\thispagestyle{empty}
\vspace*{-0.075\textheight}
\hfill\includegraphics[width=5cm]{vulcan-logo.pdf}
\vspace{0.06\textheight}

{\large\raggedright\reportdate\par}
\vspace{0.01\textheight}

{\fontsize{28pt}{30pt}\selectfont\raggedright\textbf{\reporttitle}\par}
\vspace{0.03\textheight}

{\Large\raggedright\textit{\textbf{\reportsubtitle}}\par}
\vspace{0.04\textheight}

{\large\raggedright\reportauthors\par}
\vspace{0.04\textheight}

{\normalsize\raggedright\url{https://github.com/VulcanLab/IPI-proxy}\par}
\vspace{0.06\textheight}

\begin{abstract}
\noindent
Web-browsing AI agents are increasingly deployed in enterprise settings under strict whitelists of approved domains, yet adversaries can still influence them by embedding hidden instructions in the HTML pages those domains serve. Existing red-teaming resources fall short of this scenario: prompt-injection benchmarks ship pre-built adversarial pages that whitelisted agents cannot reach, and generic LLM scanners probe the model API rather than its retrieved content. We present IPI-proxy, an open-source toolkit for red-teaming web-browsing agents against indirect prompt injection (IPI). At its core is an intercepting proxy that rewrites real HTTP responses from whitelisted domains in flight, embedding payloads drawn from a unified library of 820 deduplicated attack strings extracted from six published benchmarks (BIPIA, InjecAgent, AgentDojo, Tensor Trust, WASP, and LLMail-Inject). A YAML-driven test harness independently parameterizes the payload set, the embedding technique (HTML comment, invisible CSS, or LLM-generated semantic prose), and the HTML insertion point (6 locations from \icode{head\_meta} to \icode{script\_comment}), enabling parameter-sweep evaluation without mock pages or sandboxed environments. A companion exfiltration tracker logs successful callbacks. This paper describes the threat model, situates IPI-proxy among contemporary IPI benchmarks and red-teaming tools, and details its architecture, design decisions, and configuration interface. By bridging static benchmarks and live deployment, IPI-proxy gives AI security teams a reproducible substrate for measuring and hardening web-browsing agents against indirect prompt injection on the same retrieval surface attackers exploit in production.
\end{abstract}

\medskip\noindent\rule{\linewidth}{0.4pt}\medskip


\section{Background}
\label{sec:background}

Large-language-model (LLM) agents now routinely retrieve and reason over external content: web pages fetched through browser tools, documents loaded from corporate file shares, search results, and outputs from Model Context Protocol (MCP) servers~\cite{mcp-spec-2024}. This retrieval-augmented capability is the foundation of every modern agent product, from research assistants to autonomous developer agents. The same property that makes retrieval powerful, however, places untrusted third-party text into the same context window the model treats as authoritative instruction, collapsing the boundary between data and command. From this collapse emerges an attack class that does not exist in classical software: \emph{indirect prompt injection} (IPI), in which an adversary embeds instructions in third-party content that the agent will later read, and the LLM follows those instructions as though they were issued by the legitimate user. Two factors compound this risk in practice. First, tool-using agents let a single injected instruction hijack a chain of tool calls, exfiltrate session state, or pivot to internal systems without leaving the authorized execution surface. Second, the enterprise whitelists meant to contain agents block naive lures but not IPI itself: every wiki page, comment field, and third-party widget on an approved domain remains a viable injection point the moment an attacker can write to one. Both factors converged in the recent EchoLeak~\cite{reddy2025echoleak} zero-click exploit against Microsoft 365 Copilot, in which a single crafted email caused the production agent to exfiltrate enterprise data through legitimate auto-fetched image and Markdown link channels.

The security community has recognized this threat in its guidance. OWASP's Top~10 for Large Language Model Applications 2025~\cite{owasp-llm-2025} explicitly treats indirect prompt injection as part of LLM01: Prompt Injection. In OWASP's Top~10 for Agentic Applications 2026~\cite{owasp-agentic-2026}, \emph{Agent Goal Hijack} (ASI01) and \emph{Tool Misuse and Exploitation} (ASI02) are broader agentic risk categories, but both identify IPI as a recurring attack path (for example, through malicious content embedded in webpages, documents, external communications, or other agent-consumed data). Mitigating IPI is therefore not a peripheral hardening step, but a prerequisite for the safe deployment of tool-using, web-browsing agents.


\section{Related Work}
\label{sec:related}

The threat outlined above motivates us to study how IPI is currently understood, measured, and mitigated. This section summarizes the state of the art research and tools along four axes: (i)~attacks and threat models, (ii)~benchmarks and datasets, (iii)~defense techniques, and (iv)~general-purpose red-teaming tools.

\subsection{IPI Attacks and Threat Models}
Greshake et al.~\cite{greshake2023prompt} introduced and named the IPI threat model: rather than receiving an adversarial prompt directly from the user, the LLM retrieves third-party content (web pages, documents, emails) into which an attacker has planted instructions, and the model executes those instructions as though they were legitimate user requests. They demonstrated working exploits against deployed systems including Bing Chat and code-completion plugins and introduced the now-standard taxonomy of injection delivery channels (passive, active, hidden, and user-driven). Perez and Ribeiro~\cite{perez2022ignore} contributed an early empirical catalog of effective seed payloads such as the ``ignore previous instructions'' family. Liu et al.\ then operationalized end-to-end attacks with HouYi~\cite{liu2023houyi}, a black-box pipeline that infers an application's prompt template, generates context-aware payloads, and successfully compromised 31 of 36 commercial LLM-integrated applications, providing the first concrete evidence that injection in production systems is high-yield rather than incidental. Liu et al.~\cite{liu2024formalizing} subsequently unified attacks and defenses in a single formal framework and conducted the first systematic head-to-head evaluation of five attack families against ten defenses across ten LLMs, establishing the de facto evaluation baseline that BIPIA~\cite{yi2023bipia}, InjecAgent~\cite{zhan2024injecagent}, and AgentDojo~\cite{debenedetti2024agentdojo} all extend (Section~\ref{sec:benchmarks}). Most recently, EchoLeak~\cite{reddy2025echoleak} contributed a chained-bypass methodology demonstrating that even production-grade injection classifiers can be defeated end-to-end when the payload is smuggled through downstream rendering and link-resolution paths, showing that single-layer defenses are insufficient against well-staged in-the-wild attacks.

\subsection{Benchmarks and Datasets}
\label{sec:benchmarks}
Several large-scale datasets enumerate IPI payloads at increasing levels of realism. BIPIA~\cite{yi2023bipia} introduced the first IPI-specific benchmark, containing 250 attacker objectives across five scenarios and accompanying defenses. InjecAgent~\cite{zhan2024injecagent} ported the threat model to tool-integrated agents with 1,054 test cases over 17 user tools and 62 attacker tools. AgentDojo~\cite{debenedetti2024agentdojo} provides a dynamic environment with 97 realistic agent tasks and 629 paired security test cases that drive the agent through a tool-call loop. WASP~\cite{evtimov2025wasp} targets autonomous web UI agents specifically and shows that even frontier models achieve partial attacker success in a majority of trials. LLMail-Inject~\cite{abdelnabi2025llmailinject} is a recent adaptive challenge dataset comprising 208,095 unique attack submissions from 839 participants against an email-assistant. Tensor Trust~\cite{toyer2024tensortrust} contributes more than half a million crowd-sourced injection and defense prompts gathered through an online game. IPI-proxy treats all six of these resources as upstream payload sources rather than testing frameworks; in particular, AgentDojo's payload strings are reused, but its execution harness is not. These resources are evaluation suites, not deployable testing infrastructure: their adversarial pages are designed to be loaded by the agent under test, which presupposes that the agent can reach attacker-hosted URLs.

\subsection{Defence Techniques}
Defenses proposed against IPI fall into three broad categories. \emph{Input transformation} approaches such as Microsoft Spotlighting~\cite{hines2024spotlighting} delimit, datamark, or encode untrusted text so the LLM treats it as data rather than instruction. \emph{Training-time} approaches reshape the model itself: StruQ~\cite{chen2024struq} fine-tunes the model on structured queries that separate prompt from data, and SecAlign~\cite{chen2024secalign} uses preference optimization to reward injection-robust completions. \emph{Architectural} approaches confine the privileged capability surface around the LLM. CaMeL~\cite{debenedetti2025camel} introduces a capability-based control/data-flow layer that derives provable guarantees on AgentDojo, while Beurer-Kellner et al.~\cite{beurerkellner2025designpatterns} systematize dual-LLM, plan-then-execute, and quarantined-data patterns. These defenses are complementary to IPI-proxy, which is a testing toolkit and not itself a defense; rather, IPI-proxy provides an evaluation substrate against which defensive proposals can be exercised.

\subsection{Red-Teaming and Detection Tools}
On the practitioner side, several open-source frameworks provide generalised LLM red-teaming. Microsoft PyRIT~\cite{pyrit2024} orchestrates single- and multi-turn attack strategies with a cross-domain prompt-injection orchestrator; NVIDIA garak~\cite{garak2024} ships hundreds of probes across prompt injection, jailbreaking, and data-leakage classes; and promptfoo~\cite{promptfoo2024} integrates evaluations with CI pipelines. Detection-side tools include Rebuff~\cite{rebuff2023}, an open-source multi-layer detector combining heuristics, classifier scoring, vector lookup, and canary tokens, and Lakera Guard~\cite{lakeraguard2024}, a commercial classifier service. All of these tools target the model API: payloads are delivered directly to the LLM, not to its retrieval surface. None of them intercepts and modifies the HTTP responses that the agent's browsing tool retrieves from a whitelisted upstream.


\section{Problem Statement}
\label{sec:problem}

The literature reviewed in Section~\ref{sec:related} leaves three concrete gaps for the enterprise red-teamer who must test a web-browsing agent under whitelist enforcement. \textbf{(G1)~Whitelist mismatch.} Existing benchmarks host adversarial \emph{mock} pages on attacker-controlled domains. An agent restricted to, say, \icode{*.clientcorp.com} cannot reach those pages without policy modification, which itself voids the test. \textbf{(G2)~Static rather than in-flight content.} Benchmarks freeze the adversarial content at fixture-creation time. Production attackers, by contrast, have low-bandwidth but live influence over the content of trusted domains: a comment field, a wiki edit, a third-party widget. A red-teaming methodology that mirrors this should manipulate the live response, not a frozen mock. \textbf{(G3)~Model-only scope.} General red-teaming frameworks (PyRIT, garak, promptfoo) and detection tools (Rebuff, Lakera Guard) probe the LLM API directly. They do not exercise the path by which IPI most often manifests in practice: untrusted retrieved content reaching a privileged tool-using agent.

General-purpose LLM red-teaming tools such as PyRIT~\cite{pyrit2024}, garak~\cite{garak2024}, and promptfoo~\cite{promptfoo2024}, and detection libraries such as Rebuff~\cite{rebuff2023}, include direct-prompt-injection probes but are not IPI-specialized and operate at the model-API surface; Section~\ref{sec:related} discusses them in their own right. WASP~\cite{evtimov2025wasp} is the closest IPI-targeted neighbor but pursues a different objective: it scores autonomous \emph{web UI} agents (browser-driven, click-and-type interaction loops in WebArena-style mock environments) on whether injected instructions divert their UI actions. IPI-proxy targets the broader class of agents that retrieve HTML over HTTP, including headless fetchers, search-result readers, and MCP-mediated retrieval tools that never render a UI, and it tests them against \emph{live} responses from the same whitelisted domains they would visit in production. The two are complementary: WASP measures end-to-end UI-action robustness on fixed mock pages, while IPI-proxy measures retrieval-time susceptibility on live pages. Table~\ref{tab:comparison} therefore restricts itself to IPI-targeted resources and characterizes each entry by its kind, the temporality of its adversarial content, and the substrate on which that content reaches the agent under test.

\begin{table}[tbp]
\centering
\caption{Comparison of IPI-targeted resources and IPI-proxy. \emph{Type} classifies the artifact (Benchmark, Dataset, or Red-team tool); \emph{Static/Dynamic} distinguishes adversarial content that is frozen at fixture-creation time from content generated or applied at test runtime; \emph{Adversarial input source} is the substrate from which the attack content reaches the agent. \checkmark{} = supported, -- = not supported.}
\label{tab:comparison}
\small
\begin{adjustbox}{max width=\linewidth}
\begin{tabular}{l l l l c l}
\toprule
\textbf{Tool / Benchmark} & \textbf{Type} & \textbf{Static/Dynamic} & \textbf{Adversarial input source} & \textbf{Whitelist-resp.} & \textbf{Target surface} \\
\midrule
BIPIA~\cite{yi2023bipia}                       & Benchmark      & Static  & Text fixtures                      & -- & Model / tasks \\
InjecAgent~\cite{zhan2024injecagent}           & Benchmark      & Static  & Mock tool returns                  & -- & Tool agent \\
AgentDojo~\cite{debenedetti2024agentdojo}      & Benchmark      & Static  & Mock tool harness                  & -- & Tool agent \\
WASP~\cite{evtimov2025wasp}                    & Benchmark      & Static  & Mock web pages                     & -- & Web UI agent \\
LLMail-Inject~\cite{abdelnabi2025llmailinject} & Dataset        & Static  & Mock email inbox                   & -- & Email agent \\
Tensor Trust~\cite{toyer2024tensortrust}       & Dataset        & Static  & Crowdsourced (mostly direct PI)    & -- & Model \\
\addlinespace
\rowcolor{vulcanyellow!30}
\textbf{IPI-proxy (this work)} & \textbf{Red-team tool} & \textbf{Dynamic} & \textbf{Live HTTP responses} & \checkmark & \textbf{Browsing agent} \\
\bottomrule
\end{tabular}
\end{adjustbox}
\end{table}

Among IPI-specific resources, IPI-proxy is the only entry that pairs a Dynamic test mode with a live-HTTP adversarial input source and whitelist-respecting execution against a browsing-agent surface. It complements, rather than replaces, the benchmarks above: the payloads it injects are drawn from them.


\section{Key Idea: IPI-proxy}
\label{sec:keyidea}

The gaps identified in Section~\ref{sec:problem} share a common cause: existing tools assume the tester can dictate \emph{where} the agent goes. IPI-proxy inverts this assumption. Rather than asking the agent to visit attacker-controlled pages, it modifies the responses returned by the pages the agent is already permitted to visit. The agent never leaves its approved-domain whitelist; the proxy sits on the network path and rewrites HTML in transit. This single architectural choice closes G1 (whitelist mismatch) and G2 (static content) by construction and supports G3 (browsing-agent scope) by exercising the retrieval surface end-to-end.

Three design tenets follow from this inversion:

\begin{enumerate}
    \item \emph{Whitelist-respecting man-in-the-middle.} The proxy exposes a standard HTTP/HTTPS interception point built on mitmproxy~\cite{mitmproxy}. The agent's egress configuration points at the proxy; from the agent's perspective, it is browsing the same whitelisted domains it always browses. Payloads are inserted into responses that pass through, never into requests, and the upstream domain is never tampered with.
    \item \emph{Unified multi-source payload library.} A single deduplicated JSONL file (\icode{payloads/unified.jsonl}) holds 820 attack strings extracted from BIPIA, InjecAgent, AgentDojo, Tensor Trust, WASP, and LLMail-Inject under a uniform schema (\icode{id}, \icode{payload}, \icode{attack\_type}, \icode{domain\_context}, \icode{source\_benchmark}, \icode{severity}). Cross-benchmark filtering and rotation are configurable per test session, so a single run can mix payloads from disparate sources or isolate a specific attack class such as \icode{data\_exfil}.
    \item \emph{Decoupled embedding, insertion, and payload axes.} Three orthogonal parameters characterize an injected attack: which payload string is used, how the string is wrapped to evade trivial human inspection (HTML comment, invisible CSS, or LLM-generated semantic prose), and where in the HTML document the wrapped block is inserted (one of six insertion points). Decoupling these axes turns a single payload into a small parameter sweep, which in turn makes the success surface of a given agent-defense configuration legible.
\end{enumerate}

These tenets define IPI-proxy as a \emph{red-teaming substrate} rather than a single attack: the proxy itself is small, but the cross-product of payload~$\times$~embedding~$\times$~insertion produces a rich evaluation regime that subsumes the static use of any individual benchmark.


\section{Architecture Design}
\label{sec:architecture}

This section refines the key idea into an executable system. We describe the high-level component layout, the per-request data flow, the configuration surface exposed to the tester, and the design choices we deliberately rejected.

\subsection{System Overview}
\label{sec:arch-overview}

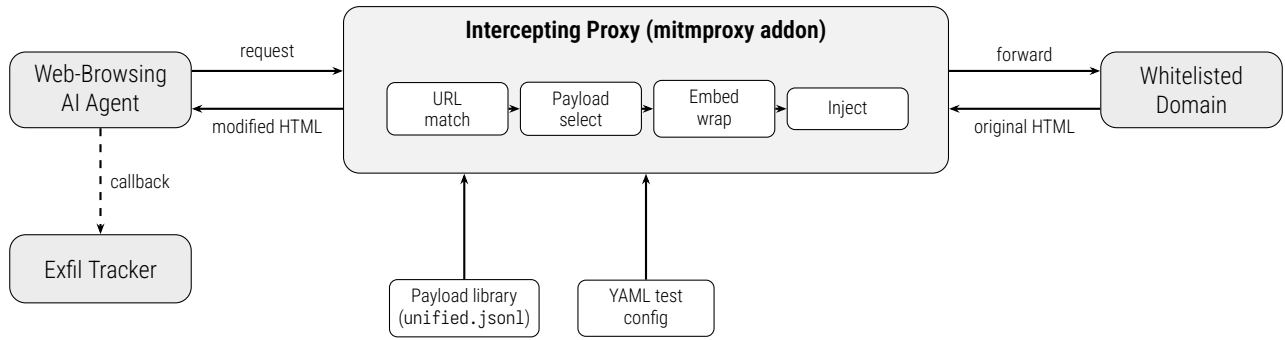
\begin{figure}[tbp]
\centering
\begin{tikzpicture}[
    box/.style={draw, rounded corners=6pt, minimum width=2.4cm, minimum height=1.0cm, align=center, font=\small, fill=gray!10},
    inner/.style={draw, rounded corners=3pt, minimum width=1.6cm, minimum height=0.55cm, align=center, font=\scriptsize, fill=white},
    src/.style={draw, rounded corners=3pt, minimum width=1.8cm, minimum height=0.6cm, align=center, font=\scriptsize, fill=white},
    arr/.style={-{Stealth[length=4pt]}, thick},
    lbl/.style={font=\scriptsize, midway, align=center}
]

\node[box, fill=gray!15] (agent) {Web-Browsing\\AI Agent};

\node[box, right=2.0cm of agent, minimum width=8.0cm, minimum height=2.2cm] (proxy) {};
\node[font=\small\bfseries, anchor=north] at ([yshift=-0.05cm]proxy.north) {Intercepting Proxy (mitmproxy addon)};

\node[inner] (s1) at ([xshift=-2.625cm,yshift=-0.25cm]proxy.center) {URL\\match};
\node[inner, right=0.15cm of s1] (s2) {Payload\\select};
\node[inner, right=0.15cm of s2] (s3) {Embed\\wrap};
\node[inner, right=0.15cm of s3] (s4) {Inject};

\draw[arr] (s1) -- (s2);
\draw[arr] (s2) -- (s3);
\draw[arr] (s3) -- (s4);

\node[box, right=2.0cm of proxy, fill=gray!15] (domain) {Whitelisted\\Domain};

\draw[arr] ([yshift=0.25cm]agent.east) -- node[lbl, above] {request} ([yshift=0.25cm]proxy.west);
\draw[arr] ([yshift=-0.25cm]proxy.west) -- node[lbl, below] {modified HTML} ([yshift=-0.25cm]agent.east);
\draw[arr] ([yshift=0.25cm]proxy.east) -- node[lbl, above] {forward} ([yshift=0.25cm]domain.west);
\draw[arr] ([yshift=-0.25cm]domain.west) -- node[lbl, below] {original HTML} ([yshift=-0.25cm]proxy.east);

\node[box, below=1.4cm of agent, fill=gray!15, minimum height=0.9cm] (tracker) {Exfil Tracker};

\node[src, below=1.4cm of proxy.south, anchor=north] (cfg) {YAML test\\config};
\node[src, left=0.5cm of cfg] (lib) {Payload library\\(\texttt{unified.jsonl})};

\draw[arr] (lib.north) -- (lib.north |- proxy.south);
\draw[arr] (cfg.north) -- (cfg.north |- proxy.south);

\draw[arr, dashed] (agent.south) -- node[lbl, right] {callback} (tracker.north);

\end{tikzpicture}
\caption{IPI-proxy architecture. The agent's browsing traffic is routed through a mitmproxy addon, which (1)~matches the destination URL against the test target, (2)~selects a payload from the unified library subject to YAML filters and rotation policy, (3)~wraps the payload in one of three embedding techniques, and (4)~injects the wrapped block at the configured HTML insertion point. The original whitelisted domain is never modified. Successful callbacks induced by the injected instruction are recorded by a separate exfiltration tracker.}
\label{fig:arch}
\end{figure}

IPI-proxy is structured around five components: a mitmproxy addon, an HTML injector, a set of embedding templates, the unified payload library, and an exfiltration tracker. Figure~\ref{fig:arch} depicts how they interact.

\subsection{Components}
\label{sec:arch-components}

\paragraph{Proxy Addon (\icode{proxy/addon.py}).}
The addon registers a \icode{response} handler with mitmproxy. On each response it consults the test configuration to decide whether the destination URL matches the target pattern, whether the response is HTML, whether the status code permits modification, and whether the per-session injection budget has been reached. When all checks pass it invokes payload selection, embedding, and injection in sequence, then writes the modified body back into the response.

\paragraph{Payload Library and Selector (\icode{payloads/unified.jsonl}, \icode{proxy/config.py}).}
At configure time the addon loads the unified JSONL library and applies YAML-defined filters on \icode{attack\_type}, \icode{source\_benchmark}, and \icode{severity}. Filtering happens once per session for performance. Rotation is sequential by default but may be set to random; an optional \icode{payloads\_per\_session} cap prevents a single test from saturating its target with thousands of injections.

\paragraph{Embedding Templates (\icode{templates/}).}
Three wrapping techniques are provided. \emph{HTML comment} (\icode{html\_comment}) places the payload inside an \icode{<!-- ... -->} block; comments are invisible to human readers but are read by LLM-based agents that consume raw HTML. \emph{Invisible CSS} (\icode{invisible\_css}) wraps the payload in a \icode{<div>} styled with \icode{display:none; left:-9999px; font-size:0}, hiding it from a rendered browser viewport while leaving it in the DOM. \emph{Semantic embedding} (\icode{semantic}) integrates the payload into a plausible prose paragraph; a deterministic template provides a default editorial-note framing, and an opt-in LLM mode (\icode{USE\_LLM=true}) calls Claude to weave the verbatim instruction into a more idiomatic paragraph.

\paragraph{HTML Injector (\icode{proxy/injector.py}).}
Given a wrapped block, the injector inserts it at one of six locations, summarized in Table~\ref{tab:insertion}. Insertion is implemented with anchored regular expressions over the response body; positions that do not exist in the document fall back transparently to a coarser anchor, ensuring that the injection succeeds on minimally structured pages.

\begin{table}[tbp]
\centering
\caption{HTML insertion points supported by IPI-proxy.}
\label{tab:insertion}
\small
\begin{tabular}{l l l}
\toprule
\textbf{Point} & \textbf{Anchor} & \textbf{Fallback} \\
\midrule
\icode{head\_meta}      & before \icode{</head>}                        & --       \\
\icode{body\_top}       & after \icode{<body...>}                       & --       \\
\icode{body\_inline}    & after first \icode{</p>}                      & \icode{body\_top}  \\
\icode{sidebar}         & before \icode{</aside>} or \icode{</nav>}     & \icode{footer}     \\
\icode{footer}          & before \icode{</body>}                        & --       \\
\icode{script\_comment} & inside first \icode{<script>} as JS comment   & append   \\
\bottomrule
\end{tabular}
\end{table}

\paragraph{Exfiltration Tracker (\icode{tracker/server.py}).}
Many IPI payloads instruct the agent to issue a follow-up callback that exfiltrates session state. The tracker is a small FastAPI service that records every request to its \icode{/track} endpoint as a JSON line, providing a ground-truth signal of successful exfiltration without requiring instrumentation of the agent under test.

\subsection{Per-Request Data Flow}
\label{sec:arch-flow}

The end-to-end path through these components for a single intercepted response is: (1)~the agent issues an HTTP request to a whitelisted URL; (2)~mitmproxy forwards the request unchanged to the upstream domain; (3)~the upstream returns the original HTML; (4)~the addon's URL matcher checks the response against the configured \icode{url\_pattern} (a glob, or a list of globs, or the wildcard \icode{*}); (5)~the payload selector returns the next entry from the filtered, rotated library; (6)~the embedding template wraps the raw payload string; (7)~the injector splices the wrapped block into the chosen insertion point; (8)~the modified response is returned to the agent. If the agent subsequently follows the injected instruction and emits a callback to the tracker URL, step~(9) logs the callback and the test is marked as a confirmed successful exfiltration.

\subsection{Configuration Interface}
\label{sec:arch-config}

A single YAML file (Figure~\ref{lst:config}) parameterizes all of the above. The four blocks (\icode{target}, \icode{injection}, \icode{verification}, \icode{rotation}) are independent: changing the embedding does not require touching the URL pattern, and rotating the payload set does not affect the tracker URL. Defaults are conservative; in particular, \icode{url\_pattern: "*"} matches every URL when no test-specific pattern is given.

\begin{figure}[tbp]
\centering
\begin{codebox}
\begin{lstlisting}[style=yamlstyle]
target:
  url_pattern: "*.clientcorp.com/docs/*"
injection:
  payload_filter:
    attack_type: data_exfil
    source_benchmark: [bipia, injecagent]
  embedding: html_comment   # or invisible_css | semantic | random
  insertion_point: body_inline
verification:
  exfil_tracker_url: "http://localhost:9090/track"
  timeout_seconds: 30
rotation:
  mode: sequential          # or random
  payloads_per_session: 10
\end{lstlisting}
\end{codebox}
\caption{Example test-case configuration.}
\label{lst:config}
\end{figure}

\subsection{Design Discussion}
\label{sec:arch-discussion}

A handful of design choices warrant explicit justification. \emph{Why mitmproxy rather than a browser extension?} An extension would tie the harness to a specific browser and would not catch traffic from headless or non-browser tool implementations. A network-level proxy is browser-agnostic, language-agnostic, and the same configuration tests an agent that switches between Playwright, a custom HTTP client, and an MCP-mediated retrieval tool. \emph{Why JSONL rather than a relational store?} The payload library is append-only, the per-line schema is small, and JSONL streams cleanly through extraction scripts and \icode{jq} pipelines. \emph{Why offer a deterministic semantic-embedding fallback?} Tests must be reproducible. The default template embedding is deterministic and dependency-free; the LLM mode is an explicit opt-in for evaluators who need maximally evasive prose. \emph{Why a separate exfiltration tracker rather than instrumenting the agent?} Decoupling the success oracle from the agent under test prevents the tester's instrumentation from biasing the agent's behavior, and matches the way an external attacker would observe a successful exfiltration in production. \emph{Ethical scope.} IPI-proxy is intended for authorized red-team engagements, security research, and defensive evaluation of one's own agents; like any MITM tool, deploying it against agents one does not control is out of scope.


\section{Conclusion}
\label{sec:conclusion}

We have presented IPI-proxy, an open-source toolkit for red-teaming web-browsing AI agents against indirect prompt injection. By inverting the conventional benchmark topology, IPI-proxy modifies live responses from whitelisted upstream domains rather than serving adversarial mock pages, allowing pre-deployment testing to honor the same egress restrictions as the production environment. A unified library of 820 payloads from six published benchmarks, three embedding techniques, and six HTML insertion points yield a parameter-sweep evaluation regime that no existing benchmark or red-teaming tool covers in combination. Several directions remain for future work: tighter agent-side observability so that intermediate tool calls can be correlated with successful exfiltrations; a browser-extension companion to test agents that resist explicit network reconfiguration; integration with continuous red-team pipelines so that newly published payloads are automatically swept against deployed agents; and extensions to non-HTML retrieval surfaces, including PDF, image-OCR, and JSON tool outputs, where IPI is increasingly being demonstrated.


\bibliographystyle{IEEEtran}
\bibliography{ref}

\end{document}